\documentclass[]{article}

\usepackage{color,amsmath,amssymb,amsfonts}
\usepackage{epsf,epsfig,subfigure,psfrag,graphicx,afterpage,chicago}
\usepackage{url,lscape}

\newcommand{\bfzero}{\hbox{\boldmath$0$}}

\newcommand{\bt}{{\scriptsize \bftheta}}

\newcommand{\bfI}{\hbox{\boldmath$I$}}
\newcommand{\bfM}{\hbox{\boldmath$M$}}
\newcommand{\bfG}{\hbox{\boldmath$G$}}

\newcommand{\bfy}{\hbox{\boldmath$y$}}
\newcommand{\bfp}{\hbox{\boldmath$p$}}

\newcommand{\bftheta}{\hbox{\boldmath$\theta$}}

\begin{document}

\title{Bayesian Inference for Generalized Extreme Value Distributions via Hamiltonian Monte Carlo}

\date{Dec 2014}

\author{Marcelo Hartmann$^{\rm a}$ and Ricardo S. Ehlers$^{\rm a}$
\thanks{$^{\ast}$ Corresponding author. Email: ehlers@icmc.usp.br}
\vspace{6pt}\\
$^{\rm a}${\em Universidade de S\~ao Paulo, S\~ao Carlos, Brazil}}

\maketitle

\begin{abstract}

In this paper we propose to evaluate and compare Markov chain Monte
Carlo (MCMC)
methods to estimate the parameters in a generalized extreme value
model. We employed the Bayesian 
approach using traditional Metropolis-Hastings methods,
Hamiltonian Monte Carlo (HMC) and Riemann manifold HMC (RMHMC) methods
to obtain the approximations to the 
posterior marginal distributions of interest. Applications to real
datasets of maxima illustrate illustrate how HMC can be much more efficient
computationally than traditional MCMC and simulation studies are conducted
to compare the algorithms in terms of how fast they get close
enough to the stationary distribution so as to provide good estimates with
a smaller number of iterations. 
\vskip .5cm

Key words:
Extreme value; Bayesian approach; Hamiltonian Monte Carlo; Markov chain Monte Carlo.

\end{abstract}

\section{Introduction}

Extreme Value Theory (EVT) can be seen as a branch of probability
theory which studies the stochastic behaviour of extremes associated
to a set of random variables with a common probability
distribution. In recent years, several statistical techniques capable
of better quantifying the probability of occurence of rare events have
grown in popularity, especially in areas such as Finance, Actuaries
and Environmental sciences (see for example, \citeNP{colw94},
\citeNP{colt96}). For a good review of both theory and interesting
applications of EVT the main reference is still \citeN{coles01}.

Natural phenomena like river flows, wind speed and rain are subject to
extreme values that can imply in great material and financial
losses. Financial markets where large amounts of money invested can
have an impact in the economy of a country need to have their risks of
large losses and gains quantified. In risk analysis, estimating future
losses by modelling events associated to default is of fundamental
importance. In Insurance, the potencial risk of high value claims
needs to be quantified and associated to possible catastrofic events
due to the large amount of money involved in payments.

The usual approach for the analysis of extreme data is based on the
Generalized Extreme Value (GEV) distribution which distribution
function is given by,

\begin{equation}
H(y|\mu,\sigma,\xi)=
\exp\left\{-\left(1+\xi~\dfrac{y-\mu}{\sigma}\right)_{+}^{-1/\xi}\right\},
\end{equation}
where $\mu$, $\sigma$ and $\xi$ are location, scale and shape
parameters respectively. The $+$ sign denotes the positive part of the argument.
We use the notation $Y\sim GEV(\mu,\sigma,\xi)$. The value
of the shape parameter $\xi$ defines the tail behaviour of the distribution.
If $\xi=0$ the distribution is defined for $y\in\mathbb{R}$ and
is called a Gumbel distribution (exponentially
decaying tail). If $\xi>0$
the distribution is defined for values $y > \mu-\sigma/\xi$, has a
lower bound and is called a Fr\'echet 
distribution (slowly decaying tail).
If $\xi<0$ the distribution is defined for values $y < \mu-\sigma/\xi$, has an
upper bound and is called a negative Weibull distribution (upper
bounded tail). 

\begin{figure}[h]\centering
\includegraphics{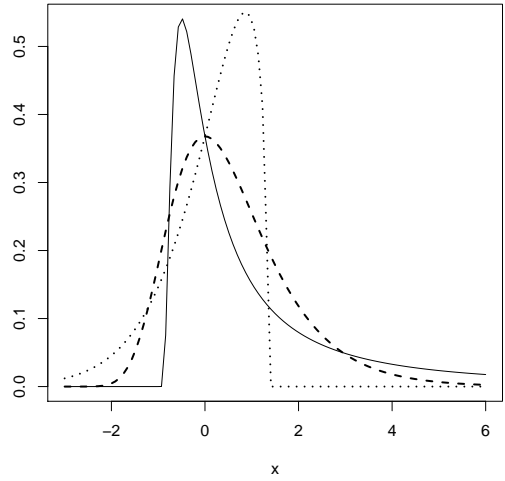}
\caption{Density functions of the GEV distribution with $\mu=0$,
  $\sigma=1$ and $\xi=1$ (full line), $\xi=0$
  (dashed line) and $\xi=-0.75$ (dotted line).}
\label{fig1}
\end{figure}

The density function of the GEV distribution is given by,
\begin{equation}
h(y|\xi,\mu,\sigma)=
\left\{\begin{array}{l}
\dfrac{1}{\sigma}\left(1+\xi~\dfrac{y-\mu}{\sigma}\right)^{-1/\xi-1}
\exp\left\{-\left(1+\xi~\dfrac{y-\mu}{\sigma}\right)^{-1/\xi}\right\},
~\xi\ne 0\\\\
\dfrac{1}{\sigma}
\exp\left\{-\left(\dfrac{y-\mu}{\sigma}\right)-
\exp\left(-\dfrac{y-\mu}{\sigma}\right)\right\}, ~\xi=0.
\end{array}
\right.
\end{equation}
which is illustrated in Figure \ref{fig1} for $\mu=0$, $\sigma=1$
and $\xi\in\{1,0,-0.75\}$.\\

\begin{center}
  Figure \ref{fig1} about here.\\
\end{center}

Now suppose that we have observed data $\bfy=(y_1,\dots,y_n)$ and
assume that they are realizations from independent and identically
distributed random variables $Y_1,\dots,Y_n$
with $Y_i\sim GEV(\mu,\sigma,\xi)$. We wish to make inferences about
the unknown parameters $\mu$, $\sigma$ and $\xi$. The
likelihood function is given by,
\begin{equation}
p(\bfy|\xi,\mu,\sigma)=
\sigma^{-n} \prod_{i=1}^n\left[1+\xi~\dfrac{y_i-\mu}{\sigma}\right]^{-1/\xi-1}
\exp\left\{-\sum_{i=1}^n\left(1+\xi~\dfrac{y_i-\mu}{\sigma}\right)^{-1/\xi}\right\}
\end{equation}
for $\mu-\sigma/\xi > y_{(n)}$ when $\xi<0$ and for $\mu-\sigma/\xi <
y_{(1)}$ when $\xi>0$. Otherwise the likelihood function is undefined.
A Bayesian analysis is then carried out by assigning prior
distributions on $\mu$, $\sigma$ and $\xi$. Simulation methods, in
particular Markov chain Monte Carlo (MCMC) methods, are now routinely employed
to produce a sample of simulated values from the posterior
distribution which can in turn be used to make inferences about the
parameters. In GEV models, the random walk Metropolis algorithm is
usually employed where a proposal distribution must be chosen and
tuned, for which a poor choice will considerably delay convergence
towards the posterior distribution. Our main motivation to 
investigate alternative algorithms is computational and we hope that
our findings are useful for the applied user of this class of models.

In the next section we describe an alternative algorithm to generate
these posterior samples in a much more efficient way. This is compared with the
traditional MCMC methods in Section \ref{sec:app} in terms of
computational efficiency through a real dataset and a simulation
study. In Section \ref{sec:ar} a time series ingredient is included in
the model to analyse time series of extreme values. Some final
comments are given in Section \ref{sec:conclusion}.

\section{Hamiltonian Monte Carlo}

Hamiltonian Monte Carlo (HMC) was originaly proposed by \shortciteN{duane} for
simulating molecular dynamics under the name of Hybrid Monte
Carlo. In what follows we present the HMC method in a compact form
which will be used in the context of GEV models.
The reader is referred to \citeN{nea2011} for an up to date review of
theoretical and practical aspects of Hamiltonian Monte Carlo methods.

Let $\bftheta\in\mathbb{R}^d$ denote a $d$-dimensional vector of
parameters, $\pi(\bftheta)$ denote the posterior density of $\bftheta$
and $\bfp\in\mathbb{R}^d$ denote a vector of auxiliary parameters
independent of $\bftheta$ and distributed as 
$\bfp\sim N(\bfzero,\bfM)$. If $\bftheta$ is interpreted as the position of a
particle 
and $-\log\pi(\bftheta)$ describes its potential energy while
$\bfp$ is the momentum with kinetic energy $\bfp'\bfM^{-1}\bfp/2$ then
the total energy of a closed system is the Hamiltonian function,

$$
H(\bftheta,\bfp)=-\mathcal{L}(\bftheta)+\bfp'\bfM^{-1}\bfp/2.
$$
where $\mathcal{L}(\bftheta)=\log\pi(\bftheta)$.\\

\noindent The (unormalized) joint density of $(\bftheta,\bfp)$ is then given by,
\begin{eqnarray*}
 f(\bftheta,\bfp) 
 \propto \pi(\bftheta) \exp(-\bfp'\bfM^{-1}\bfp/2)
 \propto \exp[-H(\bftheta,\bfp)].  
\end{eqnarray*}

For continuous time $t$, the deterministic evolution of a particle
that keeps the total energy constant is given by the Hamiltonian dynamics equations,
\begin{eqnarray*}
\frac{\partial\bftheta}{\partial t} &=&
\frac{\partial H(\bftheta,\bfp)}{\partial\bfp}=\bfM^{-1}\bfp\\
\frac{\partial\bfp}{\partial t} &=&
-\frac{\partial H(\bftheta,\bfp)}{\partial\bftheta}=\nabla_{\bt}\mathcal{L}(\bftheta).
\end{eqnarray*}
where $\nabla_{\bt}\mathcal{L}(\bftheta)$ is the gradient of
$\mathcal{L}(\bftheta)$ with respect to $\bftheta$. So, the idea is
that introducing the auxiliary variables $\bfp$ and using the
gradients will lead to a more efficient exploration of the parameter space.\\

\noindent However these differential equations cannot be solved
analytically and numerical methods are required. One such method is
the St\"ormer-Verlet (or Leapfrog) numerical integrator
(\citeNP{leimr04}) which discretizes the Hamiltonian dynamics as the
following steps,
\begin{eqnarray*}
\bfp^{(\tau+\epsilon/2)} &=& 
\bfp^{(\tau)} + \frac{\epsilon}{2}\nabla_{\bt}\mathcal{L}(\bftheta^{(\tau)})\\
\bftheta^{(\tau+\epsilon)} &=& 
\bftheta^{(\tau)} + \epsilon\bfM^{-1}\bfp^{(\tau+\epsilon/2)}\\
\bfp^{(\tau+\epsilon)} &=& 
\bfp^{(\tau+\epsilon/2)} + \frac{\epsilon}{2}\nabla_{\bt}\mathcal{L}(\bftheta^{(\tau+\epsilon)})
\end{eqnarray*}
for some user specified small step-size $\epsilon>0$. After a given
number of time steps this results in a proposal $(\bftheta^*,\bfp^*)$.
In Appendix \ref{appendix} we provide details on the required expressions of
partial derivatives for HMC. 

A Metropolis acceptance probability must then be employed to correct the
error introduced by this discretization and ensure
convergence to the invariant distribution. Since the joint distribution of
$(\bftheta,\bfp)$ is our target distribution, the transition to a new
proposed value $(\bftheta^*,\bfp^*)$ is accepted with probability,
\begin{eqnarray*}
  \alpha[(\bftheta,\bfp),(\bftheta^*,\bfp^*)] 
  &=&
  \min\left[\frac{f(\bftheta^*,\bfp^*)}{f(\bftheta,\bfp)},1\right]\\\\
  &=&
  \min\left[\exp[H(\bftheta,\bfp) - H(\bftheta^*,\bfp^*)],1\right].
\end{eqnarray*}

In the distribution of the auxiliary parameters, $\bfM$ is a symmetric
positive definite mass matrix which is typically diagonal with
constant elements, i.e.\linebreak $\bfM=m\bfI_d$.
The HMC algorithm in its simplest form taking $m=1$ is given by,
\begin{enumerate}
\item Give an initial position $\bftheta^{(0)}$ and set $i=1$,
\item\label{step2} draw $\bfp^*\sim N_d(\bfzero,\bfI_d)$ and $u\sim U(0,1)$,
\item set $(\bftheta^{(I)},\bfp^{(I)})=(\bftheta^{(i-1)},\bfp^{*})$ and $H_0=H(\bftheta^{(I)},\bfp^{(I)})$,
\item repeat the St\"ormer-Verlag solution $L$ times,\vskip .2cm
  \begin{itemize}
  \item $\bfp^* = \bfp^* + \frac{\epsilon}{2}\nabla_{\bt}\mathcal{L}(\bftheta^{(i-1)})$
  \item $\bftheta^{(i-1)} = \bftheta^{(i-1)} + \epsilon\bfp^{*}$
  \item $\bfp^* = \bfp^* + \frac{\epsilon}{2}\nabla_{\bt}\mathcal{L}(\bftheta^{(i-1)})$
  \end{itemize}
\vskip .2cm
\item set $(\bftheta^{(L)},\bfp^{(L)})=(\bftheta^{(i-1)},\bfp^{*})$ and $H_1=H(\bftheta^{(L)},\bfp^{(L)})$,
\item compute
  $\alpha[(\bftheta^{(I)},\bfp^{(I)}),(\bftheta^{(L)},\bfp^{(L)})]$ = $\min[\exp(H_0-H_1),1]$,
\item set $\bftheta^{(i)}=\bftheta^{(L)}$ if
$\alpha[(\bftheta^{(I)},\bfp^{(I)}),(\bftheta^{(L)},\bfp^{(L)})] > u$
and $\bftheta^{(i)}=\bftheta^{(I)}$ otherwise.
\item set $i=i+1$ and return to step \ref{step2} until convergence.
\end{enumerate}

\noindent Since the algorithm is making use of first derivatives of
the (unormalized) log-posterior densities it tends to propose moves to
regions of higher probabilities and the chains are expected to reach
stationarity faster. Also, in order to employ this algorithm all
sampling must be done on an unconstrained space, so we need
to implement a transformation of $\bftheta$ to the real line. Then
prior distributions are assigned and derivatives are taken for the
transformed parameters.

\subsection{Riemann Manifold Hamiltonian Monte Carlo}

\citeN{giro11} developed a modification in the proposal
mechanism in which the moves are according to a
Riemann metric instead of the standard Euclidean distance.
This procedure explores geometric properties of the posterior
distribution and is referred to as Riemann manifold HMC or RMHMC.
The idea is to redefine the Hamiltonian function as,
$$
H(\bftheta,\bfp)=
-\mathcal{L}(\bftheta)+\frac{1}{2}\log|\bfG(\bftheta)| + \frac{1}{2}\bfp'\bfG(\bftheta)^{-1}\bfp.
$$
where the position dependent matrix $\bfG(\bftheta)$ adapts to the
local geometry of the posterior distribution (see also
\shortciteNP{wang-etal}). In this paper we adopt the form proposed in
\citeN{giro11} where,
\begin{eqnarray*}
\bfG(\bftheta) = 
-E\left(\frac{d^2\mathcal{L}(\bftheta)}{d\bftheta^\top\bftheta}\right)=
-E\left(\frac{d^2 \log f(\bfy|\bftheta)}{d\bftheta^\top\bftheta}\right)
-\frac{d^2\log f(\bftheta)}{d\bftheta^\top\bftheta}
\end{eqnarray*}
i.e. the expected Fisher information matrix plus the negative Hessian of the
log-prior. The Hamiltonian dynamics becomes,
\begin{eqnarray*}
\frac{\partial\bftheta}{\partial t} &=&
\frac{\partial H(\bftheta,\bfp)}{\partial\bfp}=\bfG(\bftheta)^{-1}\bfp\\
\frac{\partial p_i}{\partial t} &=&
-\frac{\partial H(\bftheta,\bfp)}{\partial\theta_i}=\nabla_{\bt_i}\mathcal{L}(\bftheta)
-\frac{1}{2} tr\left[\bfG(\bftheta)^{-1}\frac{\partial\bfG(\bftheta)}{\partial\theta_i}\right]
+\frac{1}{2}\bfp'\bfG(\bftheta)^{-1}\frac{\partial\bfG(\bftheta)}{\partial\theta_i}\bfp.
\end{eqnarray*}
and in order to simulate values in discrete time we adopt the
generalized St\"ormer-Verlet solution (\citeNP{leimr04}). Expressions
for the expected Fisher information matrix and the Hessian of the
log-prior are provided in Appendix \ref{appendix}.

\section{Applications}\label{sec:app}

\subsection{Annual Maximum Sea Levels}

\begin{figure}[h]\centering
\includegraphics{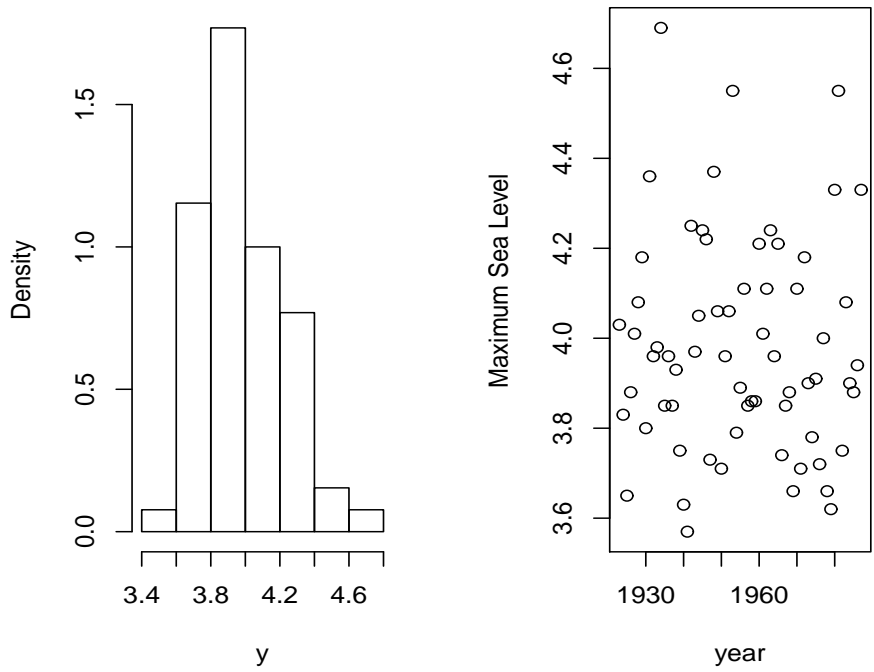}
\caption{Histogram and plots of maximum sea levels (in metres) from 1923 to 1987 at Port Pirie,
South Australia.}
\label{fig:data}
\end{figure}

This example is taken from \citeN{coles04} page 59 and refers to the
annual maximum sea levels (in metres) from 1923 to 1987 at Port Pirie,
South Australia (see Figure \ref{fig:data}). 
The objective is to fit a generalized extreme value
distribution to this data. The prior distribution adopted is a trivariate
normal on $(\mu,\log(\sigma),\xi)$ with mean vector zero and diagonal
variance covariance matrix (i.e. assuming prior independence) with prior variances
equal to 25. The complete conditional distributions are not of 
any standard form and Metropolis steps are used to yield the required
realizations from the posterior distribution.

\begin{center}
  Figure \ref{fig:data} about here.
\end{center}

\begin{figure}[p]\centering
\includegraphics{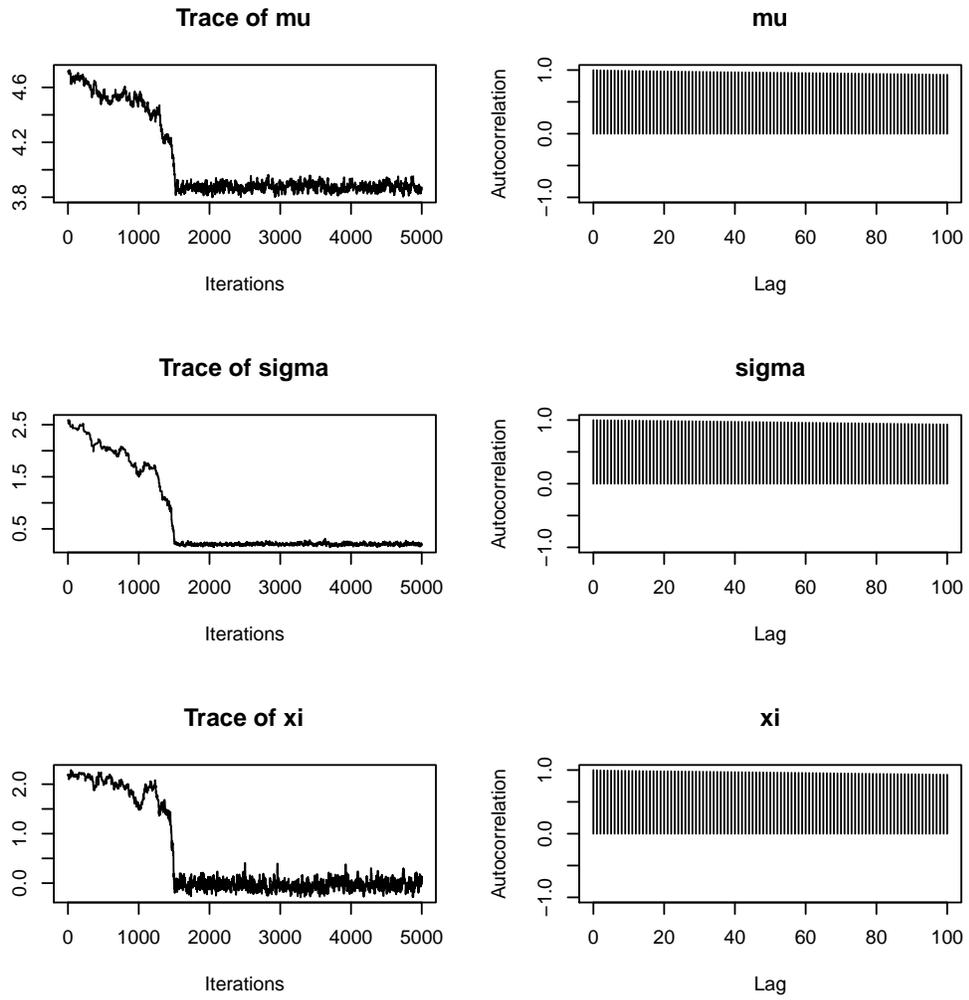}
\caption{Trace plots and autocorrelations for the parameter values
  generated using Metropolis-Hastings 
  (5000 iterations after $1000$ burn-in).}
\label{fig:mh1}
\end{figure}

For comparison purposes we also used the {\tt R} package 
{\tt evdbayes} (\citeNP{stepr06})
which is freely available from the website
{\tt http://cran.r-project.org/web/packages/evdbayes} and
provides functions for the Bayesian analysis of extreme value models
using MCMC methods. This package uses the Metropolis-Hastings algorithm.
Figure \ref{fig:mh1} shows the trace plots of the
sampled values of $\mu$, $\sigma$ and $\xi$ using the {\tt evdbayes}
package with $6000$ simulations discarding the first
$1000$ as burn-in. We note that even after discarding the
first $1000$ iterations 
the chains are far from convergence and sample autocorrelations are still high.


\begin{center}
  Figure \ref{fig:mh1} about here.
\end{center}

\begin{figure}[p]\centering
\includegraphics{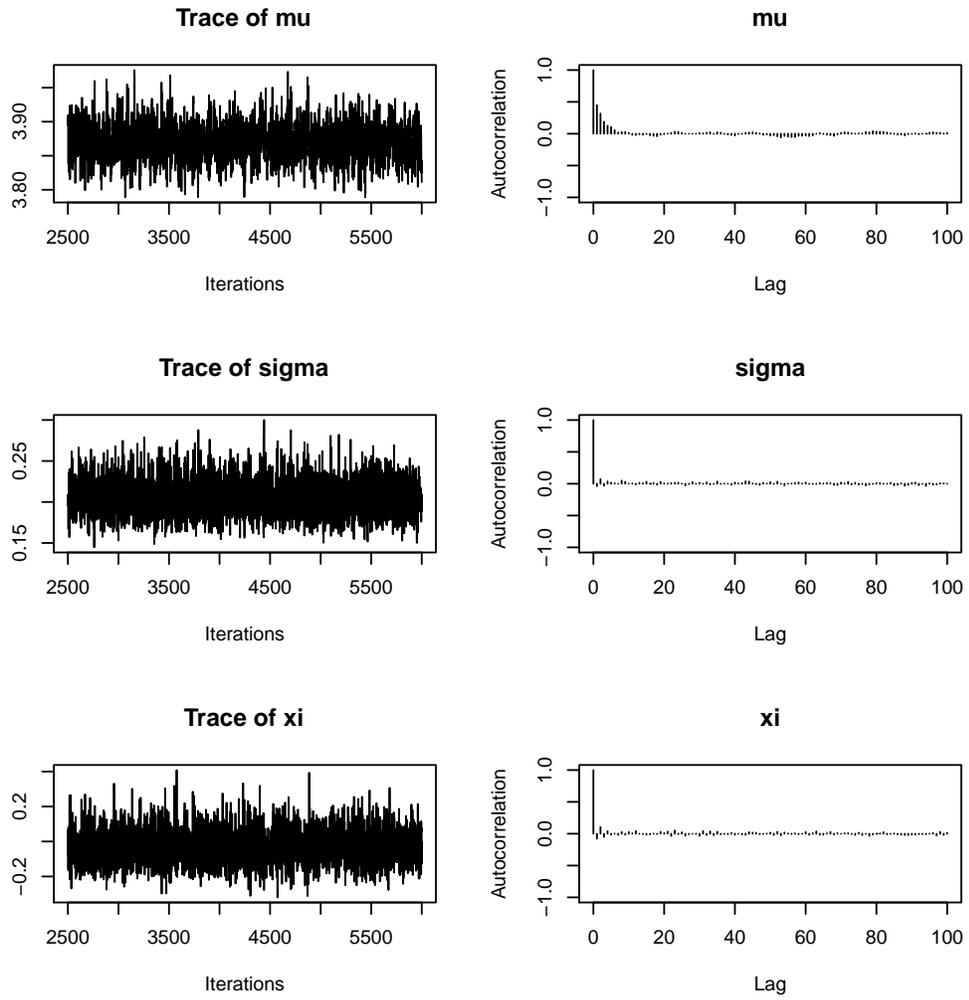}
\caption{Trace plots and autocorrelations for the parameter values
  generated using HMC 
  (5000 iterations after $1000$ burn-in).}
\label{fig:hmc}
\end{figure}

The HMC algoritm was implemented in {\tt R}. 
After some pilot tunning the parameter $\epsilon$ was taken as 0.12
and the St\"ormer-Verlet solution was replicated 27 times. The results
appear in Figure \ref{fig:hmc} which shows the trace plots of sampled values of
$\mu$, $\sigma$ and $\xi$ using HMC. 
We note that the HMC algorithm had an acceptance rate around 0.95 and
reachs a stationary regime much faster than the
Metropolis-Hastings. Besides, there is practically no autocorrelation
in the output chains.

In order to compare the relative efficiency of these methods we
calculate the effective sample size (ESS) 
using the posterior samples for each parameter. This measure is
defined as $ESS=N/(1+2\sum_k\gamma(k)$ where $N$ is the number of
posterior samples and $\gamma(k)$ are the monotone lag $k$ sample
autocorrelations (\citeNP{gey92}). 
It can thus be interpreted as the number of effectively independent
samples. For a fair comparison, first we discarded another 1500
iterations from the samples generated by MH and HMC algorithms. 
The ESS is easily obtained from any MCMC output using the
functionality from the {\tt R} package {\tt coda}
(\shortciteNP{plummer-etal06}) which provides tools
for output analysis and diagnostics. Table \ref{tab1} shows the
effective samples sizes for the parameters using both algorithms based
on the last 3500 iterations from which we can see a much lower degree of
autocorrelation in the HMC output.

\begin{center} 
  Table \ref{tab1} about here.
\end{center}

\subsection{A Simulation Study}

In order to evaluate and compare the performances of HMC and MH
algorithms two simulation studies were conducted for parameter
estimation in a GEV model. 
In both studies we generated $m=1000$ replications of $n=15,30,50,100$
observations from a GEV model with
parameters $\mu=2$, $\sigma=0.5$ and $\xi=-0.1$.
Location and scale
parameters are usually not too difficult to estimate but according to
\citeN{coles04} the value $\xi=-0.1$ is not common in practice as it
leads to distributions with too heavy tails. This makes the inferences
for this parameter more problematic.

Let $\hat{\theta}^{(i)}$ the estimate of a parameter $\theta$ for the
$i$-th replication, $i=1,\dots,m$. To evaluate
the estimation method, two criteria were considered: the bias and the mean square
error (mse), which are defined as,
\begin{eqnarray}
bias &=& \left\{\frac{1}{m} \sum_{i=1}^m \hat{\theta}^{(i)}\right\} - \theta,\label{eq1}\\
mse  &=& \frac{1}{m} \sum_{i=1}^m \left \{ \hat{\theta}^{(i)} - \theta  \right\}^2.\label{eq2}
\end{eqnarray}

For each replication and each sample size a GEV model was fitted using
the HMC and Metropolis algorithm (using {\tt evdbayes} package) based on
20000 iterations discarding 10000 as burn-in.
In this study the posterior modes were taken as parameter point
estimates in (\ref{eq1}) and (\ref{eq2}) since the marginal posterior
distributions are skewed. The 
results in terms of bias and mean square errors for each parameter
appear in Table \ref{ressim}. Overall, both measures are pretty small for both
algorithms although they tend to be slightly smaller for the HMC. This
was expected since after the 10000 iterations discarded the Metropolis
algorithm is as close to the invariant distribution as the HMC algorithm.

In a second experiment, we generated only 1100 samples from the
posterior distribution discarding the first 100 as burn-in. The main
objetive here is to see whether the HMC algorithm tends to get close enough
to the stationary distribution so as to provide good estimates with
such a small number of iterations. The results are shown in 
Table \ref{ressim1} from which we can see that both bias and mean square
error are still relatively small for the HMC algorithm while the
Metropolis algorithm appears to be definetely far from the stationary
distribution. Therefore, the advantage of adopting the HMC algorithm instead of
Metropolis seems clear at least in terms of speed of convergence. This
comes at a price of obtaining and evaluating first derivatives which
are really easy to obtain and code as shown in Appendix \ref{appendix}.
Finally, the computational times for each iteration were not too large in
this application after some pilot tunning for the step-size. Of course
each iteration of HMC takes more time than in the Metropolis algorithm but
this is more than compensated by the faster convergence (we need many
less iterations).

\begin{center} 
  Table \ref{ressim} about here.
\end{center}

\begin{center} 
  Table \ref{ressim1} about here.
\end{center}

\section{Modelling Time Dependence}\label{sec:ar}

In this section we extend the GEV model by allowing
the location parameter to vary across observations through an
autoregressive process of order $p$ (AR($p$)). The model is given by,
$$
Y_t = \mu + \sum_{j=1}^p\theta_j Y_{t-j} + e_t, ~t=1,\dots,n
$$
where $e_t$ are independent identically distributed random
errors distributed as $e_t\sim GEV(0,\sigma,\xi)$.
Assuming second order stationarity and restricting 
$\xi\in (-0.5,0.5)$ it follows that,
\begin{eqnarray}
E[Y_t]=\mu_{y_t} &=& \dfrac{\mu_{e_t} + \mu}{1-\sum_{j=1}^p \theta_j}, \forall t \nonumber \\
E[e_t]=\mu_{e_t} &=& - \dfrac{\sigma}{\xi} + \dfrac{\sigma}{\xi}\Gamma(1-\xi), \nonumber \\
Var[e_t]=\sigma^2_{e_t} &=& \dfrac{\sigma^{2}}{\xi^{2}}\left[\Gamma(1-2\xi)-\Gamma^{2}(1-\xi)\right].
\end{eqnarray}

The likelihood function is given by,
\begin{equation}\label{lik}
l(\mu,\bftheta,\sigma,\xi)=
\prod_{t=p+1}^n f(y_t|D_{t-1},\mu,\bftheta,\sigma,\xi)I_{\Omega_t}(y_t),
\end{equation}
where $D_{t-1}=(y_{t-1},\dots,y_{t-p})$ and
$\bftheta=(\theta_1,\dots,\theta_p)$. Denoting
$\mu_t = \mu + \sum_{j=1}^p\theta_j Y_{t-j}$ then 
$\Omega_t=\{y_t:1+\xi(y_t-\mu_t)/\sigma > 0\}$ and
$Y_t|\bfy_{-p},\mu,\bftheta,\sigma,\xi\sim GEV(\mu_t,\sigma,\xi)$.
\vskip .5cm

Prior distributions are then assigned to the parameters $\bftheta$,
$\mu$, $\sigma$ and $\xi$. These are assumed to be a priori
independent with relatively vague prior distributions defined in the
original parameter space, except for $\xi$ which is constrained to the
interval $(-0.5,0.5)$ so that both the mean and the variance of the
autoregressive process exist. In what follows, we adopt the prior
specifications $\theta_j\sim N(0,25)$, $j=1,\dots,p$, $\mu\sim N(0,25)$,
$\sigma\sim IG(0.1,0.001)$ and $\xi\sim U(-0.5,0.5)$.
\vskip .5cm

\subsection{A Simulation Study for GEV-AR Models}

In this simulation study, the main objective is to investigate the
behaviour of the HMC and 
RMHMC algorithms in terms of speed to reach the stationary
distribution. Therefore, in this experiment we performed only
600 MCMC iterations discarding the first 100 as burn-in.
We generated $m=1000$ replications of $n=60,150,300$ time series observations from
GEV-AR($p$) models with $p=1,2,3$. The artificial time series were
simulated from the following stationary models,
\begin{eqnarray*}
M_1: Y_t &=& -1 + 0.80Y_{t-1}+ e_t\\
M_2: Y_t &=& -1 + 0.90Y_{t-1}- 0.80Y_{t-2} + e_t\\
M_3: Y_t &=& -1 - 1.56Y_{t-1}- 0.55Y_{t-2} + 0.04Y_{t-3} + e_t
\end{eqnarray*}
where the error terms $e_t$ are independent and identicaly distributed as\linebreak 
$e_t\sim GEV(0,\sigma=1, \xi=0.3)$, $t=1,\dots,n$. 

For the HMC algorithm we set $\epsilon=0.006$ and repeated the
St\"ormer-Verlet solution 13 times. For the RMHMC, we used a fixed
metric given by the model information matrix evaluated at the MAP
estimate. For the $AR$-$GEV(1)$ and $AR$-$GEV(2)$
models the elements $E[Y_t^2]$ and $E[Y_t Y_{t+1}]$
are determined in closed form for all $t$. For the
$AR$-$GEV(3)$ model we used the approximation
$E[Y_t Y_{t+i}] \approx \mu_{Y_t}^2 + \widehat{C}(Y_t, Y_{t+i})$,
$i=0,1,2$, where $\widehat{C}$ is the sample covariance matrix. We set
$\epsilon = 0.15$ and repeated the St\"ormer-Verlet solution 13 times.

The simulation results are reported in Table \ref{ressim2} as bias and
mean square errors as defined in expressions (\ref{eq1}) and (\ref{eq2}).
For models of orders 1 and 2 and the three sample sizes considered the
performances in terms of bias are barely similar but these are in
general smaller for the RMHMC algorithm. This is also true for the model
of order 3 and sample sizes 60 and 150, but for samples of size 300
the HMC algorithm underestimates $\mu$ and $\sigma$ more severely and,
except for $\theta_1$, the biases are smaller for the RMHMC
algorithm. When we look at the mean square errors, the comparison is
in general more favorable to the RMHMC specially for larger sample
sizes. In particular, for the $AR$-$GEV(3)$ model the mean square
error tends to decrease (sometimes dramatically) for all sample sizes.
At this point, an explanation for the large values of mse for $\mu$
and $\sigma$ in the $AR$-$GEV(3)$ model is in order. Recall that we
comparing the performances of the two algorithms based on relatively
few MCMC iterations. So, for samples of size 300 the initial values
where probably far from regions of higher posterior probabilities and
the HMC would require more iterations while for the RMHMC these
initial values were much less influencial.

All in all, we consider that this simulation study provides empirical
evidence of a better performance of the RMHMC algorithm and we would
recommend this approach to the
applied user dealing with time series of extreme values.

\begin{center}
  Table \ref{ressim2} about here.
\end{center}

\subsection{A Real Data Application}

In this application, each observation represents the maximum annual
level of Lake Michigan, which is obtained as the highest mean monthly
level, 1860 to 1955 ($T = 96$ observations). The time series data can
be obtained from the Time Series Data Library repository at
{\tt https://datamarket.com/data/set/22p3/} \nocite{hyndman}

Based on the autocorrelation and partial autocorrelation functions of
the data we
propose a $AR$-$GEV(1)$ model for this dataset. To assess the quality
of predictions, we removed the last three observations from
estimation. The predictions are then compared with the actual data.
The RMHMC algorithm was applied with a fixed metric evaluated at the
MAP estimate to simulate values
from the posterior distribution of $(\mu,\theta,\sigma,\xi)$. After a
short pilot tunning a step-size $\epsilon=0.06$ was taken and the
St\"ormer-Verlet solution was repeated 11 times at each iteration. A
total of 21000 values were simulated discarding the first 1000 as burn-in.

Table \ref{tableAR1} shows the approximations for the marginal
posterior mean, standard deviation, mode, 
median and credible interval for the model parameters.
From Table \ref{tableAR1} we note that the estimated model is
stationary with high probability and the point estimate of $\xi$ is
about $-0.25$ with a small standard deviation thus characterizing a
distribution with moderate asymetry. Convergence of the Markov chains
was assessed by visual inspection of trace and autocorrelation plots
(not shown) and all indicated that the chains reached stationarity
relatively fast with low autocorrelations.

In the Bayesian approach, given $\bfy=(y_1,\dots,y_T)$, the $j$-steps
ahead predictions are obtained 
from the predictive density of $Y_{T+j}$ which is given by,
\begin{eqnarray*}
\pi(y_{_{T+j}}|\bfy) 
&=&
\int_{\Theta}f(y_{_{T+j}}|\mu+\theta y_{_{T+j-1}},\sigma,\xi)
\pi(\mu, \theta, \sigma, \xi|\bfy) d(\mu, \theta, \sigma, \xi) \nonumber\\
&=&
E_{\mu, \theta, \sigma,\xi|D}[f(y_{_{T+j}}|\mu + \theta y_{_{T+j-1}}, \sigma, \xi)].
\end{eqnarray*}
Here we propose to compute a point prediction $\hat{y}_{_{T+j}}$ of
$Y_{T+j}$ as a Monte Carlo approximation of 
the predictive expectation, $E[y_{_{T+j}}|\bfy] =
E[E[y_{_{T+j}}|\mu,\theta,\sigma,\xi,\bfy]]$. So, given a sample of $N$
simulated parameter values we sample values $y^{(i)}_{_{T+j}}$ given 
$\mu^{(i)},\theta^{(i)},y^{(i)}_{_{T+j-1}},\sigma^{(i)},\xi^{(i)}$,
$i=1,\dots,N$ which allow us to use the following approximation,
\begin{equation*}
\hat{y}_{_{T+j}}\approx
\dfrac{1}{N}\sum\limits_{i=1}^{N} y^{(i)}_{_{T+j}}
\end{equation*}
for $j=1,2,3$.

\begin{figure}[h]\centering
\includegraphics[angle=270]{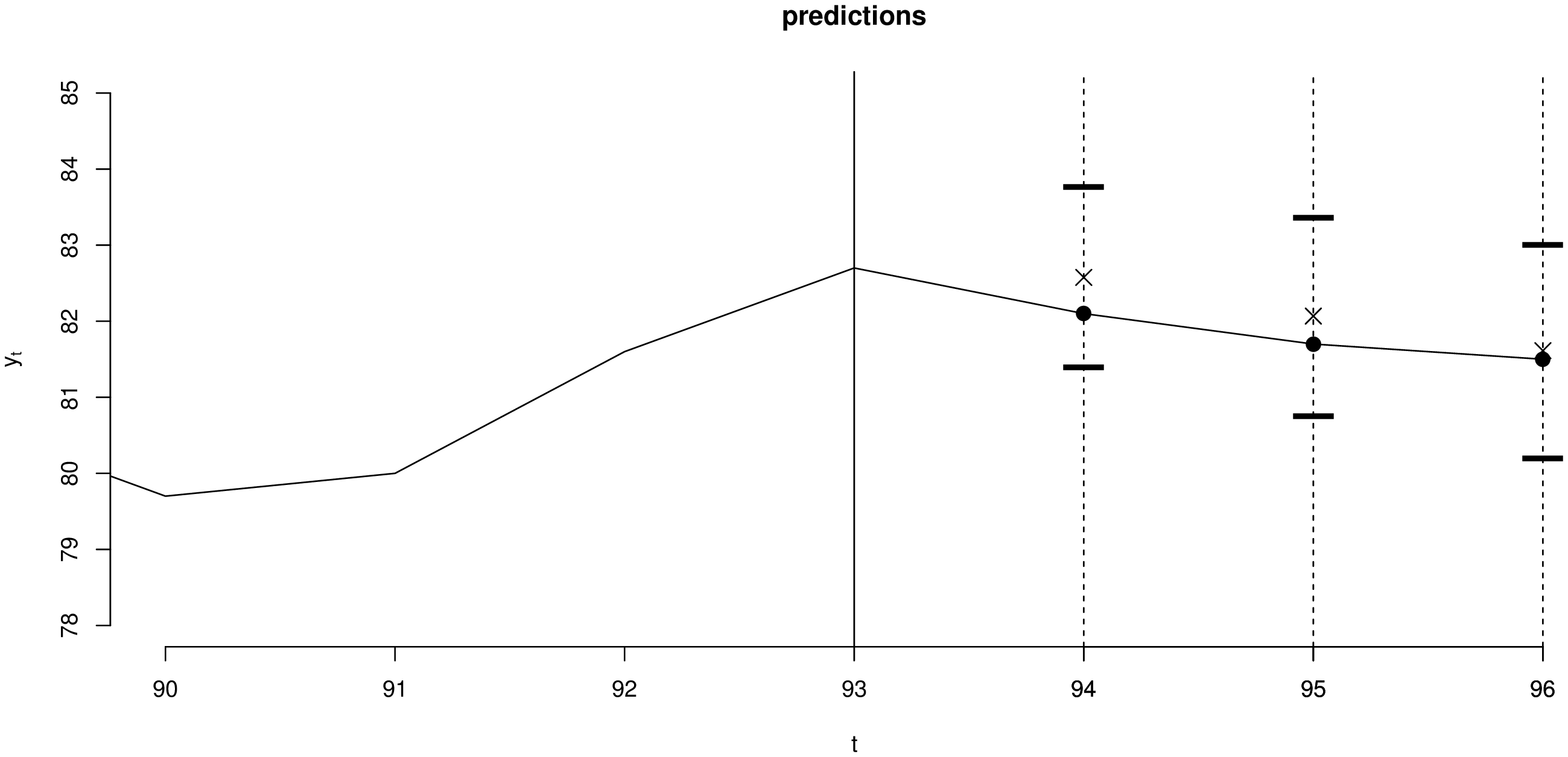}
\caption{\footnotesize Predicted values marked with an 'x' and actual
  observed values as filled circles. Horizontal bars represent the 95\%
  credible intervals.}
\label{pAR1}
\end{figure}

In Figure \ref{pAR1} we can see how the predictions behave relative to
the actual values. All observed values are within the credible
intervals of the predictive distributions which tend to follow the
time series.

\section{Conclusions}\label{sec:conclusion}

In this paper we evaluated Bayesian MCMC methods to estimate the parameters
in a generalized extreme value model both for independent and time
series data. We employed the Bayesian
approach using both traditional MCMC (Metropolis-Hastings) methods and
(Riemann manifold) Hamiltonian Monte Carlo methods to obtain the
approximations to the 
posterior marginal distributions of interest. Applications to real
datasets of maxima illustrated how (RM)HMC can be much more efficient
computationally than traditional MCMC. In a simulation study for
independent data we
noticed that parameter estimation is relatively robust to the choice
of algorithm for a large number of iterations and discarding a lot of
initial values as burn-in although bias and mean square error tend to
be slightly smaller for HMC. However, HMC was much faster to reach the
stationary distribution and this was observed by repeating the
simulations with a small number of iterations. Another simulation
study for time series data has shown that RMHMC is to be recommended
for the applied user.

As in any simulation study, our results are limited to our
particular selection of sample sizes, prior distributions and GEV
parameters. In particular, the choice $\xi=-0.1$ in Section 3.2 was intended to
compare the algorithms in a more difficult scenario in terms of
estimation (\citeNP{coles04}).
We hope that our findings are useful to the practitioners.


\section*{Acknowledgements}
The first author received financial support from CAPES - Brazil.

\appendix

\section{Appendix}\label{appendix}

In this appendix we present the expressions of gradients
needed for the implementation of HMC and RMHMC in the GEV model. In
what follows, let $z_t =  1+\xi(y_t-\mu)/\sigma$. Denoting
$\bftheta=(\mu,\sigma,\xi)$ and $L_{y|\theta}=\log f(\bfy|\bftheta)$ then,

\begin{equation*}
L_{y|\theta}=
-n\log\sigma
-\left(\frac{1}{\xi}+1\right)\sum_{t=1}^n\log\left[1+\xi~\dfrac{y_t-\mu}{\sigma}\right]
-\sum_{i=1}^n\left(1+\xi~\dfrac{y_t-\mu}{\sigma}\right)^{-1/\xi}.
\end{equation*}

\noindent The partial derivatives of this log-density with respect to the transformed
parameters $(\mu,\log(\sigma),\xi)$ are given by,
\begin{eqnarray*}
\frac{dL_{y|\theta}}{d\mu} 
&=& 
\frac{1}{\sigma}\left[(1+\xi)\sum_{t=1}^n z_t^{-1} - \sum_{t=1}^n z_t^{-1/\xi-1}\right] \\
\frac{dL_{y|\theta}}{d\delta} &=& 
-n+(1+\xi)\sum_{t=1}^n \frac{y_t-\mu}{\sigma}z_t^{-1}
-\sum_{t=1}^n \frac{y_t-\mu}{\sigma}z_t^{-1/\xi-1}\\
\frac{dL_{y|\theta}}{d\xi} 
&=&
\sum_{t=1}^n\frac{\log z_t}{\xi^2}-
\left(\frac{1}{\xi}+1\right)\left(\frac{y_t-\mu}{\sigma}\right)z_t^{-1}+
\frac{1}{\xi}\left(\frac{y_t-\mu}{\sigma}\right)z_t^{-1/\xi-1}-
\frac{\log z_t}{\xi^2}z_t^{-1/\xi}.
\end{eqnarray*}

\noindent Now letting $L_{\theta}=\log\pi(\bftheta)$ and since the
(transformed) parameters are assumed a priori independent and normally distributed
with mean zero then,
$$
\frac{dL_{\theta}}{d\mu}= -\frac{\mu}{\tau^2_{\mu}}, \quad
\frac{dL_{\theta}}{d\delta}=-\frac{\log\sigma}{\tau^2_{\sigma}}, \quad
\frac{dL_{\theta}}{d\xi}=-\frac{\xi}{\tau^2_{\xi}}.
$$
where $\tau^2_{\mu}$, $\tau^2_{\sigma}$ and $\tau^2_{\xi}$ are the prior variances.
\vskip .5cm

For the GEV-AR model we denote
$\bftheta=(\mu,\theta_1,\dots,\theta_p,\sigma,\xi)$ and the gradient
vector for the logarithm of the likelihood function 
(\ref{lik}), is a $(p+3)\times 1$ vector which elements are,
\begin{eqnarray*}
\frac{\partial L_{y|\theta}}{\partial\mu} 
&=&
\sum_{t=p+1}^T\frac{1}{\sigma}z_{t}^{-1}\left( (1 + \xi) - z_{t}^{-1/\xi} \right) \\
\frac{\partial L_{y|\theta}}{\partial\theta_i} &=&
\sum_{t=p+1}^T\frac{1}{\sigma}z_{t}^{-1}\left( (1 + \xi) -
  z_{t}^{-1/\xi} \right)y_{t-i}, ~i=1,\dots,p\\
\frac{\partial L_{y|\theta}}{\partial\sigma} 
&=&
\sum_{t=p+1}^T (1+\xi) 
\left(
\frac{y_t-\mu_t}{\sigma^2} \right) z_{t}^{-1} - \dfrac{1}{\sigma} -
z_{t}^{-(1/\xi + 1)} \left( \dfrac{y_t-\mu_t}{\sigma^2}
\right)\\
\frac{\partial L_{y|\theta}}{\partial\xi} 
&=&
\sum_{t=p+1}^T
\dfrac{\log z_{t}}{\xi^{2}}-\left(\dfrac{1}{\xi} + 1 \right) \left(
\dfrac{y_{t} - \mu_t}{\sigma}\right) z_{t}^{-1} + \dfrac{1}{\xi}
\left( \dfrac{y_{t}-\mu_t}{\sigma} \right) z_{t}^{-(1/\xi + 1)} -
\dfrac{\log z_{t}}{\xi^2}z_{t}^{-1/\xi}.
\end{eqnarray*}

\noindent To obtain the Fisher information matrix we use the fact that
$E[g(Y_t)] = E[E[g(Y_t)|D_{t-1}]], ~\forall t$. The nonzero elements
are given by,
\begin{eqnarray*}
-E\left(\dfrac{\partial^2\ell}{\partial \mu^2} \right) &=& 
-E\left[E\left( \dfrac{\partial^2\ell}{\partial \mu_t^2}\middle|
 D_{t-1}  \right) \right] = (T-p)\dfrac{A}{\sigma^{2}} \\
-E \left( \dfrac{\partial^2 \ell}{\partial \mu \partial \theta_j}
 \right) &=& (T-p)\dfrac{A}{\sigma^{2}} E[Y_{t-j}] = \mu_{Y_t}(T-p)\dfrac{A}{\sigma^{2}}\\
-E \left(\dfrac{\partial^2 \ell}{\partial \mu \partial \sigma}
\right) &=& - E \left[ E\left( \dfrac{\partial^2 \ell}{\partial \sigma
    \partial \mu_t} \middle| D_{t-1}  \right) \right] 
 = -(T-p) \dfrac{1}{\sigma^{2}\xi}[A - \Gamma(2+\xi)] \\
-E \left( \dfrac{\partial^2 \ell}{\partial \mu \partial \xi} \right)
&=& - E \left[ E\left( \dfrac{\partial^2 \ell}{\partial \xi \partial
    \mu_t}  \middle| D_{t-1}  \right) \right] 
= - (T-p)\dfrac{1}{\sigma\xi}\left(B - \dfrac{A}{\xi} \right)\\
-E \left( \dfrac{\partial^2 \ell}{\partial \theta_i \partial \theta_j}
\right) &=& - E \left[ E\left( \dfrac{\partial^2\ell}{\partial
    \mu_t^2} Y_{t-i} Y_{t-j} \middle| D_{t-1}\right)\right]
= (T-p)\dfrac{A}{\sigma^{2}} E[Y_{t-i}Y_{t-j}] \\
-E \left( \dfrac{\partial^2 \ell}{\partial \sigma \partial \theta_j}
 \right) &=& - E \left[ E\left( \dfrac{\partial^2\ell}{\partial \sigma
     \partial \mu_t} Y_{t-j} \middle| D_{t-1}  \right) \right]\\ 
 &=& -(T-p) \dfrac{1}{\sigma^{2}\xi}[A - \Gamma(2+\xi)] E[Y_{t-j}]\\
 &=& -(T-p) \dfrac{1}{\sigma^{2}\xi}[A - \Gamma(2+\xi)] \mu_{Y_t}\\
 -E \left( \dfrac{\partial^2 \ell}{\partial \xi \partial \theta_j}
 \right) &=& - E \left[ E\left( \dfrac{\partial^2\ell}{\partial \xi
     \partial \mu_t} Y_{t-j} \middle| D_{t-1}  \right) \right]\\ 
 &=& -(T-p) \dfrac{1}{\sigma\xi}\left(B - \dfrac{A}{\xi} \right) E[Y_{t-j}] \\ 
 &=& -(T-p) \dfrac{1}{\sigma\xi}\left(B - \dfrac{A}{\xi} \right)\mu_{Y_t}\\
-E\left( \dfrac{\partial^2 \ell}{\partial \xi \partial \sigma}
\right) &=& - (T-p) \dfrac{1}{\sigma\xi^{2}}\left[1 - \gamma + \dfrac{1
    - \Gamma(2+\xi)}{\xi} - B + \dfrac{A}{\xi} \right]
\end{eqnarray*}
where $A = (1+\xi)^{2}\Gamma(1+2\xi)$, 
$B = \Gamma(2+\xi)[\psi(1+\xi) + (1+\xi)\xi^{-1}]$, 
$\Gamma(\cdot)$ is the gamma function, $\psi(\cdot)$ is the digamma
function and $\gamma$ is the Euler's constant ($\cong 0.577215$).

\clearpage
\setcounter{section}{0}

\begin{table}[ht]
\begin{center}
\caption{Effective sample sizes (ESS) for each parameter using Metropolis-Hastings (MH) and Hamiltonian Monte Carlo (HMC) algorithms.}
\label{tab1}
\begin{tabular}{rrrr}
  \hline
 & $\mu$ & $\sigma$ & $\xi$ \\ 
  \hline
MH & 238.94 & 325.45 & 279.86 \\ 
  HMC & 994.11 & 2613.72 & 3427.73 \\ 
   \hline
\end{tabular}
\end{center}
\end{table}
\clearpage

\renewcommand{\arraystretch}{1.5}

\begin{table}\centering
\caption{Bias and mean squared error, based 1000 replications, for each parameter
  of the GEV distribution using Metropolis-Hastings (MH) and
  Hamiltonian Monte Carlo (HMC) algorithms.
  20000 iterations discarding 10000 as burn-in.}
\label{ressim}
\vskip .5cm
\begin{tabular}{cccccc}
\hline 
   & & \multicolumn{2}{c}{HMC} & \multicolumn{2}{c}{MH}\\
\cline{3-6} 
 $n$ & & bias & MSE & bias & MSE \\
\hline
15 & $\mu$    & -0.0008 & 0.0255 & -0.0028 & 0.0250 \\ 
   & $\sigma$ & -0.0119 & 0.0135 & -0.0121 & 0.0130 \\ 
   & $\xi$    & -0.0352 & 0.0737 & -0.0364 & 0.0727 \\
\cline{2-6}
30 & $\mu$    &  0.0000 & 0.0107 & -0.0005 & 0.0108 \\ 
   & $\sigma$ & -0.0098 & 0.0057 & -0.0084 & 0.0058 \\ 
   & $\xi$    & -0.0090 & 0.0248 & -0.0114 & 0.0256 \\ 
\cline{2-6}
50 & $\mu$    & -0.0059 & 0.0079 & -0.0045 & 0.0063 \\
   & $\sigma$ &  0.0026 & 0.0336 & -0.0028 & 0.0034 \\
   & $\xi$    & -0.0124 & 0.0149 & -0.0108 & 0.0127 \\
\cline{2-6}  
100& $\mu$    & -0.0012 & 0.0053 & -0.0010 & 0.0033 \\ 
   & $\sigma$ &  0.0022 & 0.0017 & -0.0023 & 0.0016 \\
   & $\xi$    & -0.0050 & 0.0058 & -0.0041 & 0.0053 \\
\hline
\end{tabular}
\end{table}

\clearpage

\begin{table}\centering
\caption{Bias and mean squared error, based 1000 replications, for each parameter
  of the GEV distribution using Metropolis-Hastings (MH) and
  Hamiltonian Monte Carlo (HMC) algorithms.
  1100 iterations discarding 100 as burn-in.}
\label{ressim1}
\vskip .5cm
\begin{tabular}{cccccc}
\hline 
   & & \multicolumn{2}{c}{HMC} & \multicolumn{2}{c}{MH}\\
\cline{3-6}
 $n$ & & bias & MSE & bias & MSE \\
\hline
15 & $\mu$    &  0.5169 & 0.5196 & -1.7424 & 6.0973 \\ 
   & $\sigma$ &  0.4572 & 1.5135 &  5.0180 & 51.007 \\
   & $\xi$    & -0.0681 & 0.1867 & -1.0650 & 2.5525 \\
\cline{2-6} 
30 & $\mu$    & -0.2183 & 0.3943 & -2.3592 & 8.4136 \\
   & $\sigma$ &  0.3655 & 1.0837 &  7.0782 & 78.279 \\ 
   & $\xi$    & -0.0651 & 0.0965 & -1.4178 & 3.3511 \\
\cline{2-6}
50 & $\mu$    & -0.2202 & 0.3505 & -2.6573 & 9.8133 \\ 
   & $\sigma$ &  0.3362 & 0.8232 &  8.5333 & 103.20 \\ 
   & $\xi$    & -0.0582 & 0.0542 & -1.5587 & 3.9191 \\  
\cline{2-6}  
100& $\mu$    & -0.4297 & 0.6297 & -3.2037 & 12.541 \\
   & $\sigma$ &  0.6450 & 1.7392 &  10.203 & 138.04 \\ 
   & $\xi$    & -0.0793 & 0.1241 & -1.7940 & 4.3145 \\ 
\hline
\end{tabular}
\end{table}

\clearpage

\setlength{\tabcolsep}{1mm}

\begin{landscape}
{\footnotesize
\begin{table}\centering
\caption{Bias and mean squared error, based 1000 replications, for each parameter
  of the GEV-AR model using Hamiltonian Monte Carlo (HMC) and Riemann
  manifold HMC algorithms. 600 iterations discarding 100 as burn-in.}
\label{ressim2}
\vskip .5cm
\begin{tabular}{cccccccccccccc}
\hline 
& AR-GEV$(p)$ & \multicolumn{4}{c}{$\mathbf{1}$} & \multicolumn{4}{c}{$\mathbf{2}$} & \multicolumn{4}{c}{$\mathbf{3}$} \\
\cline{1-14}
& & \multicolumn{2}{c}{HMC} & \multicolumn{2}{c}{RMHMC} & \multicolumn{2}{c}{HMC} & \multicolumn{2}{c}{RMHMC} & \multicolumn{2}{c}{HMC} & \multicolumn{2}{c}{RMHMC} \\ 
 $n$ & & bias & mse & bias & mse & bias & mse & bias & mse & bias & mse & bias & mse \\
\hline
60  &$\mu$        &-0.0236 & 0.5600 &-0.0292 &0.7966 &-0.0175 &0.3382 & 0.0086 & 0.0829 &-0.0323 &1.1502 &-0.0339 &1.2462\\ 
    &$\sigma$     &-0.0238 & 0.5701 &-0.0269 &0.6677 &-0.0039 &0.0173 &-0.0124 & 0.1506 &-0.0124 &0.1701 &-0.0129 &0.1817\\ 
    &$\xi$        & 0.0276 & 0.7667 & 0.0322 &0.9849 & 0.0111 &0.1300 & 0.0291 & 0.8020 & 0.0290 &0.9294 & 0.0283 &0.8730\\
    &$\theta_{1}$  & 0.0233 & 0.5459 & 0.0173 &0.2704 & 0.0058 &0.0381 & 0.0076 & 0.0570 & 0.0021 &0.0048 &-0.0038 &0.0159\\
    &$\theta_{2}$  &        &        &        &       &-0.0050 &0.0278 &-0.0058 & 0.0332 & 0.0121 &0.1611 & 0.0033 &0.0119\\
    &$\theta_{3}$  &        &        &        &       &        &       &        &        & 0.0095 &0.0995 & 0.0085 &0.0787\\ 
\cline{2-14} 
150 &$\mu$        & 0.0018 &0.0035 & 0.0007 &0.0005 &-0.0100 &0.1115 &-0.0077 &0.0668 &-0.0953 &10.006 &-0.0376 &1.5560 \\ 
    &$\sigma$     &-0.0242 &0.5899 &-0.0135 &0.1843 &-0.0007 &0.0005 &-0.0011 &0.0015 &-0.0863 &8.2037 &-0.0323 &1.1485 \\ 
    &$\xi$        & 0.0053 &0.0282 &-0.0016 &0.0027 & 0.0022 &0.0053 & 0.0025 &0.0071 & 0.0369 &1.5019 & 0.0170 &0.3194 \\
    &$\theta_{1}$  & 0.0144 &0.2095 & 0.0009 &0.0926 & 0.0005 &0.0002 & 0.0008 &0.0006 &-0.0082 &0.0745 &-0.0087 &0.0815 \\
    &$\theta_{2}$  &        &       &        &       &-0.0014 &0.0023 &-0.0018 &0.0038 &-0.0060 &0.0406 &-0.0055 &0.0334 \\
    &$\theta_{3}$  &        &       &        &       &        &       &        &       &-0.0004 &0.0002 & 0.0020 &0.0047 \\
\cline{2-14}
300 &$\mu$        &-0.0009 &0.0008 & -0.0002 & 0.0054 & -0.0051 & 0.0286 & -0.0048 & 0.0257 & -0.3205 & 106.06 & -0.0400 & 1.6533 \\ 
    &$\sigma$     &-0.0293 &0.8555 & -0.0058 & 0.0344 & -0.0073 & 0.0588 & -0.0053 & 0.0315 & -0.3208 & 106.26 & -0.0444 & 2.0343 \\ 
    &$\xi$        & 0.0225 &0.5082 & -0.0007 & 0.0005 &  0.0005 & 0.0003 &  0.0000 & 0.0000 & -0.0471 & 2.2938 & -0.0136 & 0.1923 \\
    &$\theta_{1}$  & 0.0232 &0.5391 &  0.0053 & 0.0289 &  0.0012 & 0.0017 &  0.0015 & 0.0027 & -0.0046 & 2.1750 & -0.0221 & 0.5036 \\
    &$\theta_{2}$  &        &       &         &        & -0.0011 & 0.0014 & -0.0016 & 0.0028 & -0.0471 & 2.2938 & -0.0136 & 0.1923 \\
    &$\theta_{3}$  &        &       &         &        &         &        &         &        & -0.0136 & 0.1924 &  0.0007 & 0.0005 \\ 
\hline
\end{tabular}
\end{table}
}
\end{landscape}

\clearpage

\begin{table}
\centering
\begin{tabular}{ccccc}
\hline 
N = 20000 &                  $\mu$ & $\theta$ & $\sigma$ & $\xi$\\
\hline
$\widehat{E[.|D]}$           & 5.929 & 0.923 & 0.692 & -0.258  \\
$\widehat{DP[.|D]}$          & 3.350 & 0.041 & 0.055 &  0.058  \\
$\widehat{\mathrm{Moda}}$    & 6.369 & 0.922 & 0.687 & -0.261  \\
$\widehat{\mathrm{Mediana}}$ & 5.945 & 0.923 & 0.689 & -0.259  \\
$IC \ 95\%$                  & [0.443, 11.437] & [0.856, 0.991] & [0.609, 0.790] & [-0.351, -0.160]\\ 
\hline
\end{tabular}
\caption{\footnotesize Posterior mean, standard deviation, mode,
  median and credible interval.}
\label{tableAR1}
\end{table}

\clearpage

\setkeys{Gin}{width=3in,height=3in}

\clearpage
\setkeys{Gin}{width=5in,height=4.5in}

\clearpage
\setkeys{Gin}{width=5.5in,height=5.5in}

\clearpage
\setkeys{Gin}{width=5.5in,height=5.5in}

\clearpage

\begin{thebibliography}{99}

\bibitem[\protect\citeauthoryear{Coles}{Coles}{2001}]{coles01}
Coles, S.~G. (2001).
\newblock {\em Extreme Value Theory and Applications}.
\newblock Kluver Academic Publishers.

\bibitem[\protect\citeauthoryear{Coles}{Coles}{2004}]{coles04}
Coles, S.~G. (2004).
\newblock {\em An Introduction to Statistical Modelling of Extreme Values}.
\newblock Springer Series in Statistics.

\bibitem[\protect\citeauthoryear{Coles and Tawn}{Coles and Tawn}{1996}]{colt96}
Coles, S.~G. and J.~A. Tawn (1996).
\newblock A {B}ayesian analysis of extreme rainfall data.
\newblock {\em Applied Statistics\/}~{\em 45\/}(4), 463--478.

\bibitem[\protect\citeauthoryear{Coles and Walshaw}{Coles and
  Walshaw}{1994}]{colw94}
Coles, S.~G. and D.~Walshaw (1994).
\newblock Directional modelling of extreme wind speeds.
\newblock {\em Applied Statistics\/}~{\em 43}, 139--157.

\bibitem[\protect\citeauthoryear{Duane, Kennedy, Pendleton, and Roweth}{Duane
  et~al.}{1987}]{duane}
Duane, S., A.~D. Kennedy, B.~J. Pendleton, and D.~Roweth (1987).
\newblock Hybrid {M}onte {C}arlo.
\newblock {\em Physics Letter B\/}~{\em 195\/}(2), 216--222.

\bibitem[\protect\citeauthoryear{Geyer}{Geyer}{1992}]{gey92}
Geyer, C.~J. (1992).
\newblock Practical {M}arkov chain {M}onte {C}arlo.
\newblock {\em Statistical Science\/}~{\em 7}, 473--511.

\bibitem[\protect\citeauthoryear{Girolami and Calderhead}{Girolami and
  Calderhead}{2011}]{giro11}
Girolami, M. and B.~Calderhead (2011).
\newblock Riemann manifold {L}angevin and {H}amiltonian {M}onte {C}arlo
  methods.
\newblock {\em Journal of the Royal Statistical Society B\/}~{\em 73},
  123--214.

\bibitem[\protect\citeauthoryear{Hyndman}{Hyndman}{}]{hyndman}
Hyndman, R.~J.
\newblock Time series data library.
\newblock http://data.is/TSDLdemo.
\newblock Accessed: 2014-03-30.

\bibitem[\protect\citeauthoryear{Leimkuhler and Reich}{Leimkuhler and
  Reich}{2004}]{leimr04}
Leimkuhler, B. and S.~Reich (2004).
\newblock {\em Simulating {H}amiltonian Dynamics}.
\newblock Cambridge University Press, New York.

\bibitem[\protect\citeauthoryear{Neal}{Neal}{2011}]{nea2011}
Neal, R.~M. (2011).
\newblock {MCMC} using {H}amiltonian dynamics.
\newblock In {\em Handbook of Markov chain Monte Carlo}. Boca Raton: Chapman
  and Hall-CRC Press.

\bibitem[\protect\citeauthoryear{Plummer, Best, Cowles, and Vines}{Plummer
  et~al.}{2006}]{plummer-etal06}
Plummer, M., N.~Best, K.~Cowles, and K.~Vines (2006).
\newblock {CODA}: Convergence diagnosis and output analysis for {MCMC}.
\newblock {\em R News\/}~{\em 6\/}(1), 7--11.

\bibitem[\protect\citeauthoryear{Stephenson and Ribatet}{Stephenson and
  Ribatet}{2006}]{stepr06}
Stephenson, A.~G. and M.~A. Ribatet (2006).
\newblock {\em A User's Guide to the {\tt evdbayes} Package (Version 1.1)}.

\bibitem[\protect\citeauthoryear{Wang, {Mohamed}, and {de Freitas}}{Wang
  et~al.}{2013}]{wang-etal}
Wang, Z., S.~{Mohamed}, and N.~{de Freitas} (2013).
\newblock {Adaptive {H}amiltonian and {R}iemann Manifold {M}onte {C}arlo
  Samplers}.
\newblock {\em ArXiv e-prints\/}.

\end{thebibliography}
\end{document}